# Mixing and flow transition in an optimized electrokinetic turbulent micromixer

Keyi Nan [1,*], Yanxia Shi [1,*], Tianyun Zhao[2], Xiaowei Tang[2], Yueqiang Zhu[1], Kaige Wang[1], Jintao Bai[1], and Wei Zhao[1,$]

1 State Key Laboratory of Photon-Technology in Western China Energy, International Collaborative Center on Photoelectric Technology and Nano Functional Materials, Institute of Photonics & Photon Technology, Northwest University, Xi'an 710069, China

2 School of Automation, Northwestern Polytechnical University, Xi'an 710072, China

* The authors contributed equally to this investigation

$ Correspondence: zwbayern@nwu.edu.cn (W.Z.);

**ABSTRACT:** Micromixer is a key element in lab on a chip for broad applications in the analysis and measurement of chemistry and engineering. Previous investigations reported electrokinetic (EK) turbulence could be realized in a "Y" type micromixer with a cross-sectional dimension of 100 µm order. Although the ultrafast turbulent mixing can be generated at a bulk flow Reynolds number of O(1), the micromixer has not been optimized. In this investigation, we systematically investigated the influence of electric field intensity, AC frequency, electric conductivity ratio, and channel width at the entrance on the mixing effect and transition electric Rayleigh number in the "Y" type electrokinetic turbulent micromixer. It is found the optimal mixing is realized in a 350 µm wide micromixer, under 100 kHz and $1.14 \times 10^5$ V/m AC electric field, with an electric conductivity ratio of 1:3000. Under the conditions, a maximum degree of mixedness of 0.93 can be achieved at 84 µm from the entrance and 100 ms. A further investigation of the critical electric field and the critical electric Rayleigh number indicates the most unstable condition of EK flow instability is inconsistent with that of the optimal mixing in EK turbulence. To predict the evolution of EK flow under high $Ra_e$ and guide the design of EK turbulent micromixer, it is necessary to apply a computational turbulence model, instead of linear instability analysis.

In recent years, micromixer has been developing rapidly and widely used in many disciplines, e.g. biology [1-7], chemistry [8-14], energy industry [15,16] and medicine [17-20], for chemical detection and analysis. Relies on the functional mechanism, micromixer is usually divided into two categories: passive micromixer and active micromixer. Passive micromixer is a kind of mixer that enhances mixing depending on specifically designed shape or structures of the microchannel. By folding and stretching fluids, the spatial distance of mixtures can be reduced to a small scale where diffusion becomes dominant in the mixing. In the operation of a passive micromixer, no additional energy is needed to generate flow disturbance [21]. The energy consumption and cost of passive micromixer are relatively low [21], however, the function of passive micromixer is normally specific and cannot realize diverse manipulation of fluids. Active micromixer is a kind of micromixer that uses external physical excitation, such as acoustic excitation [22] or electrokinetic flow [23,24] etc, to generate flow disturbance and enhance mixing.

Conventionally, the dimensions of micromixer are below sub-millimeter and the characteristic velocity of the fluid is on the orders of millimeter per second or smaller. Thus, the Reynolds number $Re = \rho U d/\mu$ (where $\rho$ is the density of the fluid, $U$ is the characteristic velocity of the fluid, $d$ is a characteristic length and $\mu$ is dynamic viscosity.) is normally on the orders of unity or below. Consequently, it is considered that the flow in microfluidics can only be laminar or chaotic, not turbulent flow which is usually considered to be generated at a high Reynolds number. The relatively low mixing efficiency is the major obstacle in many applications, such as protein folding kinetics [25], reduction of by-products in chemical reactions [26], genetic diagnosis and detection [27,28], where high mixing efficiency is essential. Therefore, improving the efficiency of micromixers has always been the goal of researchers.

In 2014, Wang et al. [29] designed a novel electrokinetic micromixer that can realize turbulent-like flow under a bulk flow Reynolds number of O(1). Prior to the study, only chaotic flow has been reported to enhance mixing due to the low Reynolds number [30]. In 2016, the same group characterized the velocity [31] and concentration fields [32] of micro electrokinetic turbulence in detail with single-point measurement methods. From the temporal measurement, they surprisingly observed the celebrated Kolmogorov -5/3 law and Obukhov-Corrsin -5/3 law in both the velocity and concentration spectra respectively. They also found the velocity structure function and concentration structure function exhibited self-similarity and high intermittency. All these features were only observed in turbulence with ultrahigh Re. In 2017, Zhao et al. [33] studied the mixing in the cross-section of the EK micromixer from two aspects of large-scale and small-scale. They found that rapid mixing can be achieved not only at the centerline, but also near the bottom wall. They hypothesized there existed a large-scale secondary flow caused by the imbalance of AC electroosmotic flow (ACEOF) near the top and bottom walls. Then, in 2020, Nan et al. [34] observed the 3D flow field using micro-particle image velocimetry (μPIV), and verified the existence of a large-scale swirling flow which can enhance entrainment in the μEK turbulent micromixer.

Although Wang and his collaborators have made a series of investigations on many aspects of EK turbulent micromixer [35], there still exists too many control parameters that could affect the transport and cascade process of kinetic energy and scalar variance [36,37], e.g. the AC frequency, width of the microchannel, conductivity ratio and the corresponding electric field intensity. The type of micromixer has far from being optimized. Therefore, in the present work, the concentration field of EK micromixer was characterized by flow visualization based on laser induced fluorescence (LIF), to explore the optimal mixing parameters of the EK turbulent micromixer, including the AC frequency, width of the microchannel, conductivity ratio and the corresponding electric field intensity. Besides, we also characterize the relationship between the parameters with transition electric Rayleigh number where the flow becomes unstable. The discrepancy between the AC frequency at the most unstable mode and that of the optimal mixing is clearly elucidated.

**Experimental setup.** In this investigation, the scalar EK turbulence in a micromixer was studied by an inverted epi-fluorescence microscopy system. The schematic of the system is shown in Figure 1a. It consists of a 473 nm laser (CNI, MW-BL-473, China), SCMOS camera (PCO edge. 4.2LT, Germany), inverted fluorescence microscope (Nexcope, NIB900, China), objective lens (Leica, 10X NA 0.32, Germany), dichroism mirror and other optical elements. The laser passes through the microscope, reflected on the dichroic mirror, and irradiates the microchannel of micromixer. The fluorescent signals transmit the dichroic mirror (Semrock, Di01-R488/543/635, U.S.A, which has high transmittance in 500 ~ 528 nm and high reflectivity in 380 ~ 491 nm) and a bandpass filter (Chroma, ZET488/640m, U.S.A), then, captured by the SCMOS camera with the exposure time of 1 ms at 40.58 fps. Fluorescent sodium salt was used as fluorescent agent.

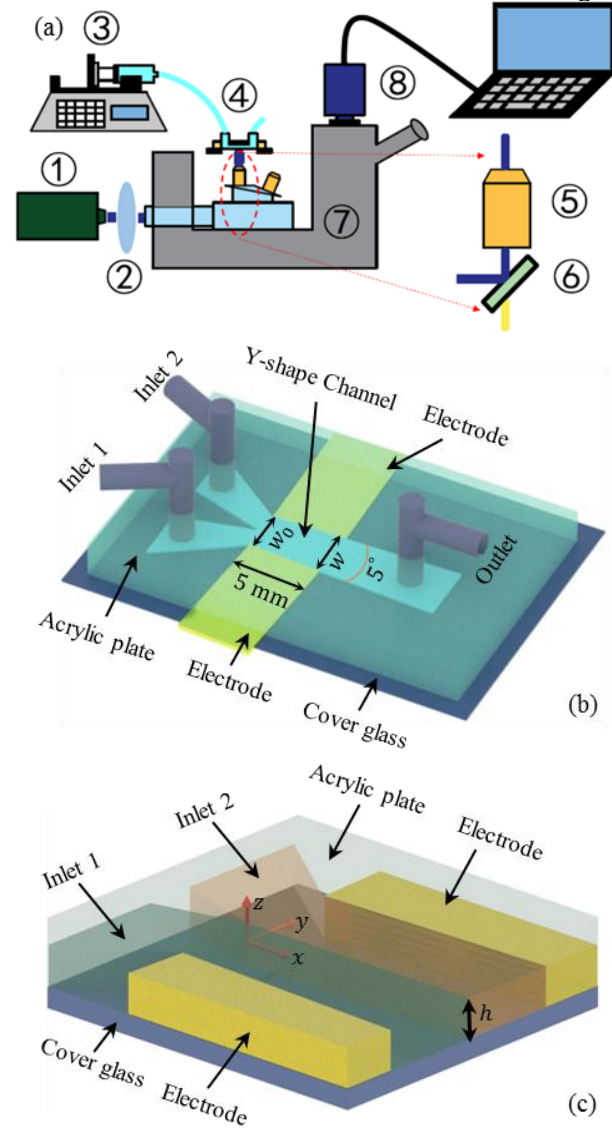

**Figure 1**. Schematic diagram of experimental system and microchannel. (a) Schematic diagram of the experimental system, includes ① 473 nm laser, ② lens set, ③ syringe pump, ④ EK micromixer, ⑤ dichroic mirror and bandpass filter, ⑥ objective lens, ⑦ inverted fluorescence microscope, ⑧ SCMOS camera. (b) Schematic diagram of the EK micromixer, and (c) the coordinate system in the EK micromixer.

Its excitation peak wavelength is 460 nm and the emission peak wavelength is 515 nm respectively.

The micromixer used in this experiment is developed according to Wang et al [29], as shown in Figure 1b. It is composed of three layers. The top layer is a 2 mm thick acrylic plate with two inlets and one outlet, the bottom layer is a 0.15 mm thick cover glass, and the middle layer is a Y-shaped channel sealed by double side tapes (3M, Maplewood, NJ, USA). The channel of the micromixer has a splitter plate with a trailing edge at the entrance, followed by a mixing chamber which has a 5° divergence angle on the side walls. Two micromixers with the same height ($h = 380$ μm) and length ($l = 10$ mm), but different widths ($w_0$) at the entrance. One has a $w_0$ of 350 μm, and the other has $w_0 = 650$ μm, respectively. The side wall of the channel is constructed by platinum electrodes contact the working fluids. Deionized water and fluorescent dye solutions were injected into the two inlets with a dual-channel syringe pump (Harvard pump 33, CA, USA) respectively. The flow rate of each stream is 3 μL/min and the conductivity ratio between the two streams is adjustable.

**Theory.** The evolution of EK flow is governed by Navier-Stokes equation with electric volume force [38], as follows

$$\rho \frac{D\boldsymbol{u}}{Dt} = -\nabla p + \mu \nabla^2 \boldsymbol{u} + F_e \quad (1)$$

where $\rho$ is fluid density and $\boldsymbol{u} = u\vec{i} + v\vec{j} + w\vec{k}$ is the instant velocity vector. $u$, $v$ and $w$ are the velocity components in streamwise ($x$), spanwise ($y$) and vertical ($z$) directions respectively as shown in Figure 1b. $\vec{i}, \vec{j}$ and $\vec{k}$ are the corresponding unit vectors in the directions. μ is the dynamic viscosity of the fluid and $D/Dt = \partial/\partial t + \boldsymbol{u} \cdot \nabla$. $F_e$ is electric body force (EBF) which can be expressed as[37,39]

$$\vec{F_e} = \rho_e \vec{E} - \frac{1}{2}(\vec{E} \cdot \vec{E})\nabla \varepsilon \quad (2)$$

with $\rho_e = -\varepsilon \vec{E} \cdot \nabla \sigma / \sigma$, where $\varepsilon$ and $\sigma$ are permittivity and conductivity of solutions. It can be seen from Eq. (2) that the EBF is closely related to the transport process of $\sigma$, which can be described by the following equation

$$\frac{D\sigma}{Dt} = D_\sigma \nabla^2 \sigma \quad (3)$$

where $D_\sigma$ is effective diffusivity of electric conductivity and $D_\sigma \approx 1.5 \times 10^{-9}\ m^2 s^{-1}$. Besides, the transport equation of fluorescent dye is

$$\frac{DC}{Dt} = D_C \nabla^2 C \quad (4)$$

where $D_C$ is the diffusivity of fluorescent dyes.

In this research, we adopt the dimensional analysis according to Zhao and Wang [37], as
$t^* = tU_e/w, x^* = x/w, y^* = y/w, z^* = z/w,$
$\vec{u}^* = \vec{u}/U_e, p^* = p/\rho U_e^2, \sigma^* = \sigma/\langle\sigma\rangle, \vec{E^*} = \vec{E}/E_w,$
$\nabla^* = w\nabla, C^* = C/\langle C \rangle \quad (5)$

where $U_e = \sqrt{2\varepsilon E_w^2(1-\beta^2)(\sigma_2-\sigma_1)/\langle\sigma\rangle\rho}$ is the characteristic velocity representing the influence of electro-inertial effect, $\sigma_1$ and $\sigma_2$ are the lower and higher electric conductivities of the two streams respectively, $\langle\sigma\rangle$ is the ensemble-averaged value of $\sigma$, $E_w = V_A/w$ is electric intensity, $V_A$ is the amplitude of the applied AC voltage. Accordingly, the dimensionless momentum and scalar transport equations can be obtained as

$$\frac{\partial \vec{u}^*}{\partial t^*} + \vec{u}^* \cdot \nabla^* \vec{u}^* = \nabla^* p^* + \frac{1}{\sqrt{Gr_e}}\nabla^{*2}\vec{u}^* - \frac{1}{2(1-\beta^2)}\left(\frac{\nabla^*\sigma^*}{\sigma^*} \cdot \vec{E^*}\right)\vec{E^*} \quad (6)$$

$$\frac{\partial \sigma^*}{\partial t^*} + \vec{u}^* \cdot \nabla^* \sigma^* = \frac{1}{\sqrt{Sc_\sigma Ra_\sigma}}\nabla^{*2}\sigma^* \quad (7)$$

$$\frac{\partial C^*}{\partial t^*} + \vec{u}^* \cdot \nabla^* C^* = \frac{1}{\sqrt{Sc_C Ra_C}}\nabla^{*2}C^* \quad (8)$$

where electric Grashof number $Gr_e = 2\varepsilon E_w^2 w^2(1-\beta^2)(\sigma_2-\sigma_1)/\rho v^2\langle\sigma\rangle$ and electric Rayleigh number $Ra_\sigma = Sc_\sigma Gr_e$, $Sc_\sigma = v/D_\sigma$ is the effective Schmidt number of ion transport, $Ra_C = Sc_C Gr_e$ is the Rayleigh number related to concentration transport accompanied with $Sc_C = v/D_C$ being Schmidt number of concentration transport, $v = \mu/\rho$ is kinematic viscosity, $\beta = 2\pi f_{AC}\varepsilon/\langle\sigma\rangle$ is a dimensionless frequency, which represents the ratio of charge relaxation time to AC electric field period. Since the two streams have equal flow rates, $\langle\sigma\rangle = (\sigma_2 + \sigma_1)/2$, the control parameter of EK turbulence — $Ra_\sigma$ can be rewritten as

$$Ra_e = 4\frac{\sigma_2-\sigma_1}{\sigma_2+\sigma_1}\varepsilon E_w^2 w^2(1-\beta^2)/\rho v D_\sigma \quad (9)$$

with $\beta = 4\pi f_{AC}\varepsilon/(\sigma_2 + \sigma_1)$. Compared with other commonly used expressions of electric Rayleigh number [29,38], Eq. (9) not only reflects the influence of the electric field intensity, the conductivity ratio between two streams and the width of micromixer, but also the influence of AC electric field frequency.

**Experimental results.** From Eqs. (1-4), it can be seen EK flow is in fact an active transport process of $\sigma$, while a passive transport process of $C$. Normally, $D_C$ and $D_\sigma$ are not the same. Accordingly, the Schmidt number related to electric conductivity and dye concentration could be different. The transport and cascade processes of $\sigma$ and $C$ are not exactly the same. However, the difference is primarily on small scales, like Batchelors' scale ($l_B = \eta/Sc_C^{1/2}$, where $\eta$ is Kolmogorov scale) and an even smaller scale $l_Z = l_K Sc_\sigma^{-5/6}$ ($l_K$ is the smallest characteristic scale for energy transport in EK turbulence) [33]. The transport of $\sigma$ can penetrate to $l_Z$, while that of C have to stop at $l_B$. Nevertheless, restricted by the spatial resolutions and sensitivity of the fluorescent microscope, it is difficult to distinguish the concentration structures on the aforementioned small scales, i.e. $l_B$ or $l_Z$. Therefore, we can assume the transport of $\sigma$ and C have the same processes in the considered scale ranges

**Mixing effects by flow visualization.** The flow visualizations of the mixing process and the concentration structures have been shown in Figure 2. When no electric field was applied, the flow is laminar and stratified, with a clear concentration interface at the center of microchannel, as shown in Figure 2a. The fluctuation of the two streams is very small. At this time, the mixing is mainly caused by molecular diffusion. The small fluctuation may be caused by the vibration of the injection pump. When the applied voltage is $V_A = 4$ V, i.e. $E_W = 1.14 \times 10^4$ V/m with $f_{AC} = 20$ kHz, the interface between the two streams is stretching, folding and turning up (Figure 2b). The flow is no longer laminar, and the vortex structure can be clearly observed. The mixing is significantly enhanced by entrainment compared with that without forcing, however, the mixing is still far from nonuniform. The vortex structures locate primarily in the central region and cannot invade the higher $\sigma$ region. When the applied $V_A$ is increased to 40 V (i.e. $E_W = 1.14 \times 10^5$ V/m) with $f_{AC} = 20$ kHz, the fluorescence intensity becomes highly uniform in the entire mixing chamber, as can be seen in Figure 2c. The fluids in the cross-section of the entire mixing chamber are well mixed and there is no obvious concentration structure, which is close to the situation of complete mixing.

Figure 2a-c clearly shows that as the $E_W$ increases from 0 V/m to $1.14 \times 10^5$ V/m, the mixing effect experiences no mixing enhancement, entrainment on large scales (nonuniform large-scale mixing), and high mixing enhancement at multiscales (uniform large-scale mixing, nonuniform and nonequilibrium small-scale mixing). Similar results can also be found in Figure 2d and e, where $f_{AC} = 100$ kHz. The difference is the mixing at $f_{AC} = 100$ kHz is visually faster than that at $f_{AC} = 20$ kHz.

The mixing effect increases with the increase of electric field intensity, which is consistent with the previous research results of Wang et al [29] and Zhao et al [33]. In the investigation, the upper limit of $V_A$ is 40 V. Further increasing $V_A$ leads to the generation of bubble near electrodes which significantly affect the mixing.

**Degree of mixedness.** To evaluate the efficiency of mixing, the degree of mixedness proposed by Zhao et al. [33] has been applied, as

$$\lambda = \frac{1}{w(x)} \int_y \lambda_{xy} dy \quad (10)$$

with the local mixing index ($\lambda_{xy}$) defined as

$$\lambda_{xy} = 1 - \frac{\sqrt{\overline{(C-\langle C \rangle_y)^2}}}{\sqrt{\langle C \rangle_y (C_0 - \langle C \rangle_y)}} \quad (11)$$

where $C_0$ is the initial concentration of the fluorescent dye $\langle C \rangle_y = \frac{1}{w(x)} \int_y \bar{C}(x,y) \, dy$ is the spatially averaged concentration along the transverse ($y$) direction of the microchannel, $\bar{C}(x,y)$ is the mean field of concentration. For a fully mixed solution, $\lambda = 1$. While for two immiscible and stratified fluids with steady flow status, $\lambda = 0$.

Figure 3 shows the degree of mixedness measured under different experimental parameters along streamwise direction. Obviously, when unforced, the mixing mainly depends on the diffusion of molecules. The degree of mixedness of the two streams is below 0.2 even at $x^* = 2$. When a 100 kHz electric field is applied, with the increase of $E_W$, the mixing is apparently enhanced, and the degree of mixedness increases as well (Figure 3a). When $E_W = 1.14 \times 10^5$ V/m, a degree of mixedness of 0.93 can be reached at 84 μm away from the trailing edge after 100 ms. The mixing time is comparable with our earlier investigation, while the mixing length required to achieve the same degree of mixedness is reduced in this investigation.

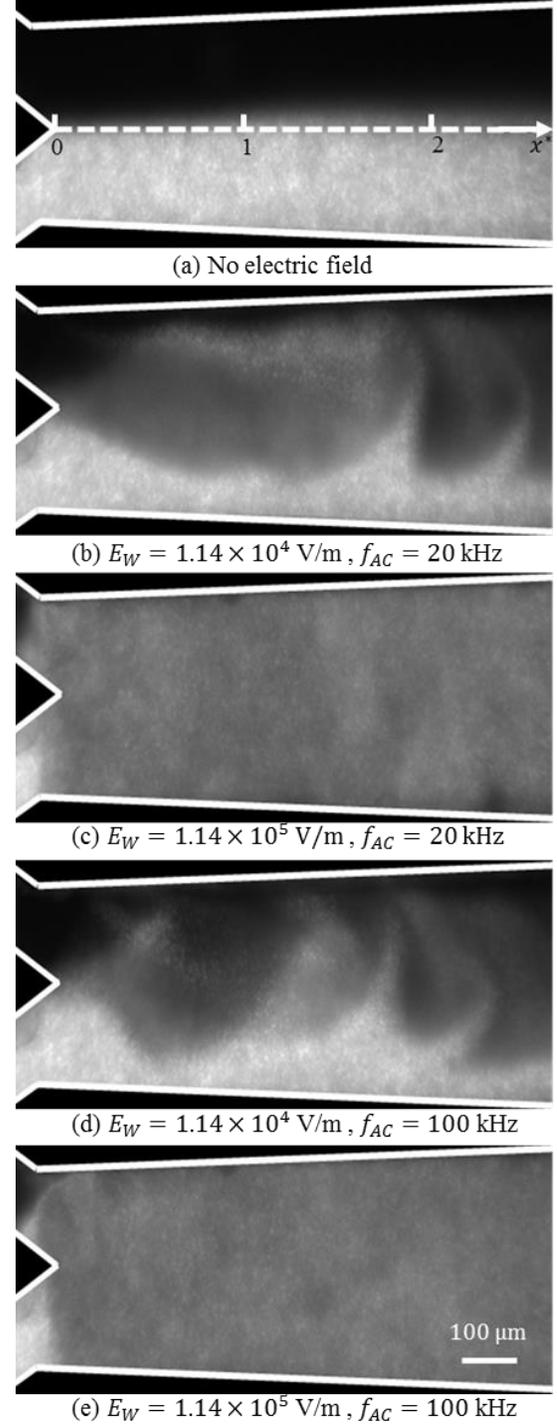

Figure 2. Mixing evaluated by flow visualization. The conductivity ratios of the two streams in the experiments are all 1:3000. The exposure time of the camera is 1 ms. Figure (a)-(e) are the fluid mixing of different voltages and frequency of AC electric field in the $w_0 = 350$ μm wide micromixer. (a) No electric field. The dark side is deionized water, while the bright side is the fluorescent dye solution. (b) $E_W = 1.14 \times 10^4$ V/m and $f_{AC} = 20$ kHz. (c) $E_W = 1.14 \times 10^5$ V/m and $f_{AC} = 20$ kHz. (d) $E_W = 1.14 \times 10^4$ V/m and $f_{AC} = 100$ kHz. (e) $E_W = 1.14 \times 10^5$ V/m and $f_{AC} = 100$ kHz.

Figure 3b shows the influence $f_{AC}$ on the mixing performance. In a broad frequency range (20 kHz to 140 kHz), the degree of mixedness achieves the maximum at $f_{AC} = 100$ kHz. Nevertheless, it should be noted, in the range of 80-140 kHz, the degree of mixedness shows negligible difference.

We also examine the influence of electric conductivity ratio ($\sigma_1:\sigma_2$) on the mixing effect, as shown in Figure 3c. It can be seen, mixing is not so fast for a large electric conductivity ratio in this investigation. There exists an optimal ratio at 1:3000, under which the mixing is the fastest.

In Figure 3d, we show the influence of $w_0$ on the degree of mixedness at equivalent $E_W$. Although the EK flow in 350 μm wide micromixer has a smaller $Ra_e$ than that in 650 μm micromixer, the former exhibits faster mixing than the latter. Compared with previous studies [33], the 350 μm wide micromixer can reach the highest degree of mixedness within the shortest distance.

From Figure 3, it can be clearly found, the curves of degree of mixedness can always be separated into two regions. One is located near the trailing edge and upstream of the critical positions ($x_c$) marked by the color dots (Figure 3a). In this region, $\lambda$ increases rapidly and in an approximately linear manner with large slopes. The fast mixing is related to the entrainment of large eddies generated at the entrance (see Figure 2), which accomplishes the large-scale mixing and initiates the multi-scale concentration structure.

At the downstream of the critical positions, another region where the mixing approaches saturation can be found. In this region, the transport of concentration structures is primarily carried out on small scales through a turbulent scalar cascade (which will be elucidated in another manuscript). With the increase of $E_W$, the dot which represents the critical position of the two regions in streamwise direction gradually moves towards the trailing edge.

**Critical position ($x_c^*$) in streamwise direction.** The critical position of the two regions in the streamwise direction determines the length of the linear region, which is important for designing a highly compact micromixer, in case an ultrahigh degree of mixedness is not required. The shorter the linear region, the smaller the micromixer could be.

The normalized critical position, say $x_c^* = x_c/w_0$, strongly depends on $Ra_e$, as shown in Figure 4. In Figure 4a, we measured the relations between $x_c^*$ and $Ra_e$, under different AC frequencies. It shows that $x_c^*$ becomes smaller with the increase of $Ra_e$ for all the applied AC frequencies. The smallest $x_c^*$ can be only 0.18. Surprisingly, almost all the data points roughly fall on a unique curve.

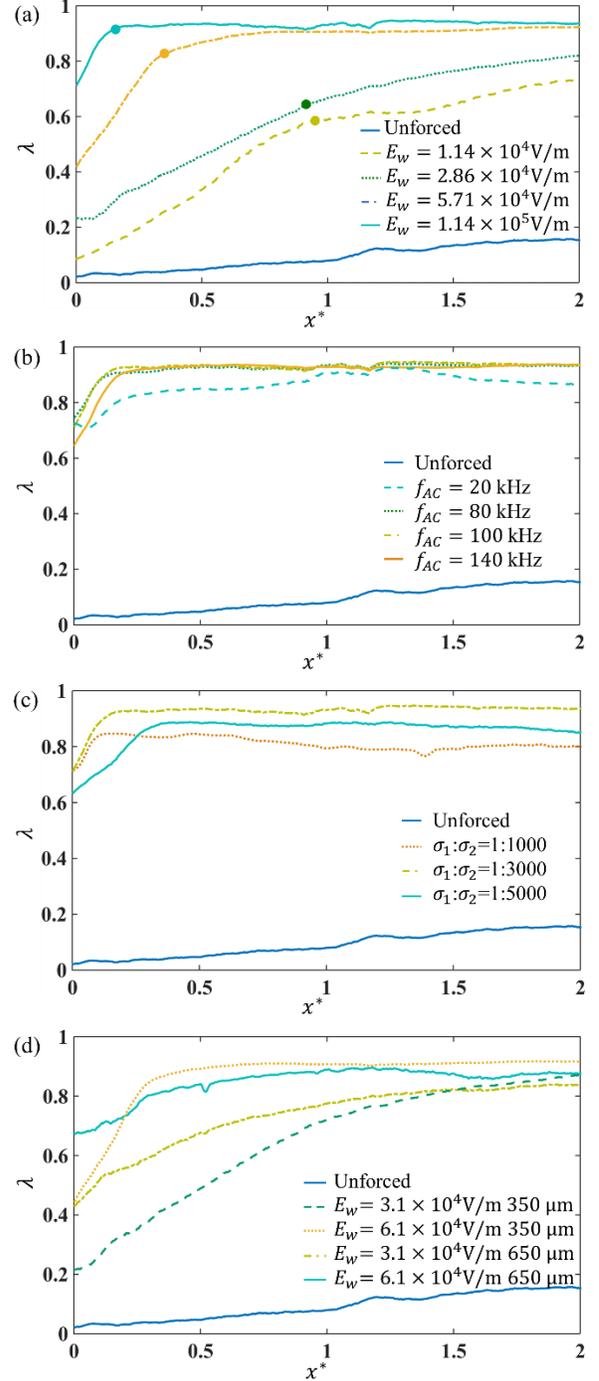

**Figure 3.** Degree of mixedness along streamwise $x$ direction. (a) $\lambda$ under different $E_W$ at $f_{AC} = 100$ kHz in the micromixer with 350 μm width. Here, the conductivity ratio is 1:3000. The dots represent the critical streamwise positions of the mixing layer from linear spreading to nonlinear saturation. (b) $\lambda$ under different $f_{AC}$ at $E_W = 1.14 \times 10^5$ V/m in the micromixer with 350 μm width. Here, the conductivity ratio is 1:3000. (c) $\lambda$ under different electric conductivity ratios at $f_{AC} = 100$ kHz and $E_W = 1.14 \times 10^5$ V/m, in the micromixer with 350 μm width. (d) $\lambda$ measured in the micromixers with different widths and $E_W$. Here, the electric conductivity ratio is 1:3000 and $f_{AC} = 100$ kHz.

Figure 4b shows the variation of $x_c^*$ versus $Ra_e$ for a variety of electric conductivity ratios in the 350 μm wide micromixer. Although the broad range of electric conductivity ratio leads to a scattering of data points, $x_c^*$ still exhibits a decreasing relationship with $Ra_e$. Similar results can also be found in Figure 4c, where the relations between $x_c^*$ and $Ra_e$ are investigated in two micromixers with different $w_0$.

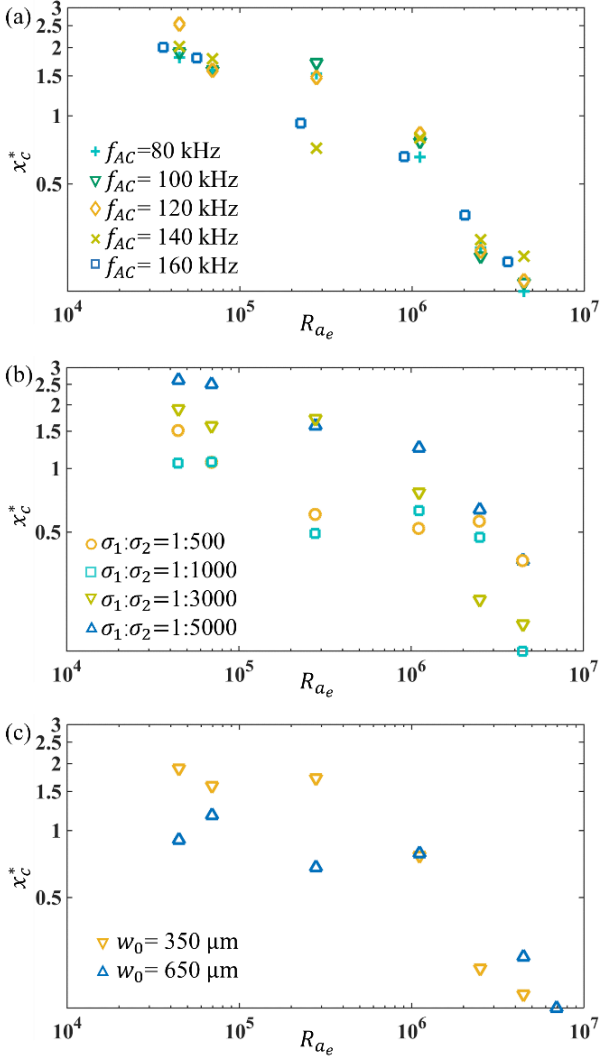

**Figure 4.** Streamwise position of the critical point. (a) $x_c^*$ vs $Ra_e$ for different AC frequencies in the micromixer with $w_0 = 350$ μm and the conductivity ratio of the two streams is 1: 3000. (b) $x_c^*$ vs $Ra_e$ according to different electric conductivity ratios, when $w_0 = 350$ μm and the frequency of the AC electric field is 100 kHz. (c) $x_c^*$ vs $Ra_e$ according to different channel widths, when the conductivity ratio of two streams is 1: 3000 and the frequency of AC electric field is 100 kHz.

We plot all the data points in Figure 5, under all AC frequencies and electric conductivity ratios. It can be seen, although the scattering exists, $x_c^*$ always exhibit approximately linear relation with $Ra_e$ in the log-log plot, for both 350 μm (Figure 5a) and 650 μm (Figure 5b) micromixers. In other words, $x_c^*$ can be approximately described by a power-law function of $Ra_e$. By nonlinear fitting, the data points of $x_c^*$ in the micromixer of $w_0 = 350$ μm can be universally expressed as

$$x_c^*(Ra_e) = 42.95 Ra_e^{-0.29} \qquad (12)$$

When in the case of $w_0 = 650$ μm, we have

$$x_c^*(Ra_e) = 15.98 Ra_e^{-0.28} \qquad (13)$$

From equations (12) and (13), the scaling exponents are -0.29 and -0.28 respectively, which are sufficiently close to each other. The result indicates with the same topology of the micromixer, even with different transverse dimensions, there roughly exists a universal expression for the length of the linear region of the EK turbulent micromixer.

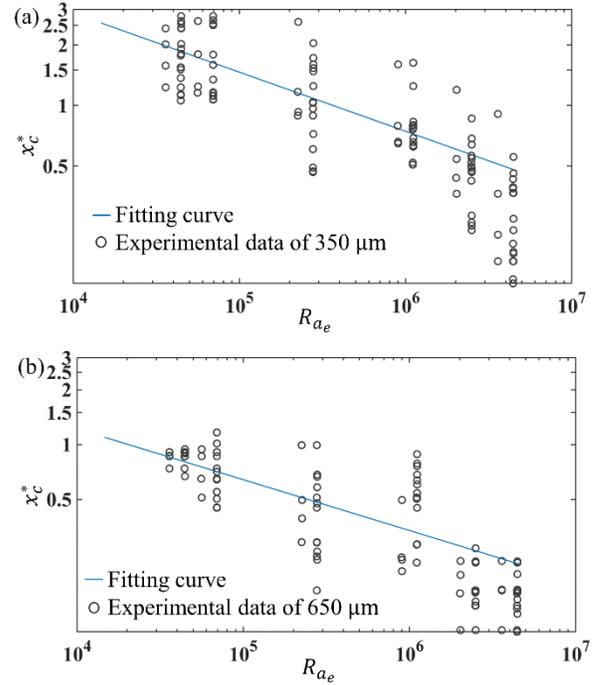

**Figure 5.** Universal relations between $x_c^*$ and $Ra_e$. (a) 350 μm wide micromixer, (b) 650 μm wide micromixer.

**Transition electric Rayleigh number.** The existence of optimal $f_{AC}$ may be arbitrarily attributed to some AC EK instability which disturbs the flow strongly. Intuitively, the smaller the threshold electric field ($E_C$) beyond which the flow becomes unsteady, the stronger the flow instability (or receptivity), which can be easier to induce unsteady flow and stronger flow mixing accordingly. This is also the major intention of most numerical investigations on EK flow instability. However, in the following section, we want to show that the $f_{AC}$ and electric conductivity ratio for the optimal mixing is different to that of the most unstable linear instability mode.

In experiments, $E_C$ is determined when the interface of two streams is initially perturbed with small fluctuations by the applied AC electric field. From the flow visualization, when $E_W < E_C$, the flow is stable and the interface between the two streams is not perturbed. While $E_W = E_C$, small and periodic perturbations emerge on the interface between the two streams. $E_C$ also represents the threshold electric field for the linear instability of electrokinetic flow to become dominant, in the schematic of sidewall electrodes with a small divergent angle.

In this section, since $E_C$ is normally small, AC voltage is directly applied through the function generator without the amplifier. The flow visualization results are shown in Figure 6. Relative to the unforced case (Figure 2a), the interface becomes slightly perturbed at $E_C$, which is dependent on $f_{AC}$, as shown in Figure 6a-f. For instance, when $f_{AC} = 1$ kHz, $E_C = 6.3 \times 10^3$ V/m. When $f_{AC} = 160$ kHz, $E_C$ reduces to $4.7 \times 10^3$ V/m. Generally, $E_C$ decreases with $f_{AC}$.

We plot the transition electric Rayleigh number $Ra_{e,c} = 4\frac{\sigma_2 - \sigma_1}{\sigma_2 + \sigma_1}\varepsilon E_C^2 w^2(1-\beta^2)/\rho\nu D_\sigma$ with different $f_{AC}$ and electric conductivity ratios in Figure 7. As can be seen from Figure 7a, all the $Ra_{e,c}$ decreases with $f_{AC}$ in an approximately linear manner. Let $Ra_{e,c} = A_1 f_{AC} + A_2$, where $A_1$ is a slope characterizing the sensitivity of $Ra_{e,c}$ on $f_{AC}$ and $A_2$ is the $Ra_{e,c}$ for DC EK flow, we can see all the $A_1$ are negative, as summarized in Table 1. Figure 7a and Table 1 indicate the AC EK flow has a smaller $Ra_{e,c}$ at larger $f_{AC}$.

Table 1. Linear fitting on the experimental data with $Ra_{e,c} = A_1 f_{AC} + A_2$.

| $\sigma_1:\sigma_2$ | $A_1$ | $A_2$ |
| --- | --- | --- |
| 1:1000 | -45.93 | $1.1034 \times 10^4$ |
| 1:3000 | -21.79 | $8.895 \times 10^3$ |
| 1:5000 | -23.72 | $7.94 \times 10^3$ |
| 1:7000 | -12.69 | $4.653 \times 10^3$ |

This is interesting. Since the smaller the $Ra_{e,c}$, the EK flow at the applied $f_{AC}$ is more unstable. The smallest $Ra_{e,c}$ should be experimentally related to the most unstable mode of EK flow instability. From Figure 7a, the smallest $Ra_{e,c}$ always appears at $f_{AC} = 200$ kHz.

In Figure 7b, the smallest $Ra_{e,c}$ always appears at $\sigma_1:\sigma_2 = 1:7000$. In contrast, in the frequency range of 60 to 200 kHz, $Ra_{e,c}$ is always the maximum at $\sigma_1:\sigma_2 = 1:3000$. In other words, it is the most difficult to perturb EK flow through AC electric field at $\sigma_1:\sigma_2 = 1:3000$.

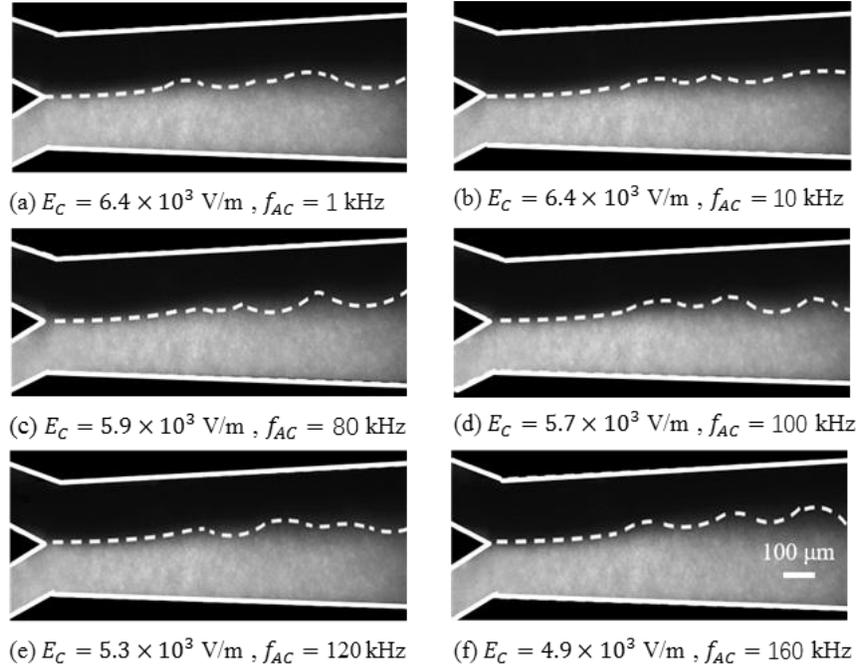

(a) $E_C = 6.4 \times 10^3$ V/m, $f_{AC} = 1$ kHz
(b) $E_C = 6.4 \times 10^3$ V/m, $f_{AC} = 10$ kHz
(c) $E_C = 5.9 \times 10^3$ V/m, $f_{AC} = 80$ kHz
(d) $E_C = 5.7 \times 10^3$ V/m, $f_{AC} = 100$ kHz
(e) $E_C = 5.3 \times 10^3$ V/m, $f_{AC} = 120$ kHz
(f) $E_C = 4.9 \times 10^3$ V/m, $f_{AC} = 160$ kHz

**Figure 6.** Perturbed interface generated at $E_C$ under different $f_{AC}$, where $w_0 = 350$ µm and $\sigma_1:\sigma_2 =$1:3000. The interfaces are highlighted by dashed lines. (a) $E_C = 6.4 \times 10^3$ V/m at $f_{AC} = 1$ kHz, (b) $E_C = 6.4 \times 10^3$ V/m at $f_{AC} = 10$ kHz, (c) $E_C = 5.9 \times 10^3$ V/m at $f_{AC} = 80$ kHz, (d) $E_C = 5.7 \times 10^3$ V/m at $f_{AC} = 100$ kHz, (e) $E_C = 5.3 \times 10^3$ V/m at $f_{AC} = 120$ kHz, (f) $E_C = 4.9 \times 10^3$ V/m at $f_{AC} = 160$ kHz.

These results indicate the optimal mixing condition, i.e. $f_{AC} = 100$ kHz and $\sigma_1:\sigma_2 = 1:3000$, is not the most unstable condition for the EK flow. There is no one-to-one correspondence between the linear instability mode of EK flow and the optimal mixing conditions in EK turbulence. This is consistent with the investigations on macroflow turbulence, and indicates a linear instability analysis is insufficient to predict the mixing related dynamic process (e.g. heat transfer, mass transfer and reaction process etc) in strongly nonlinear EK turbulence. High-efficiency numerical simulation methods, e.g direct numerical simulation, are also required in micro/nanoscale flow investigations and the corresponding development of micro/nanofluidics techniques.

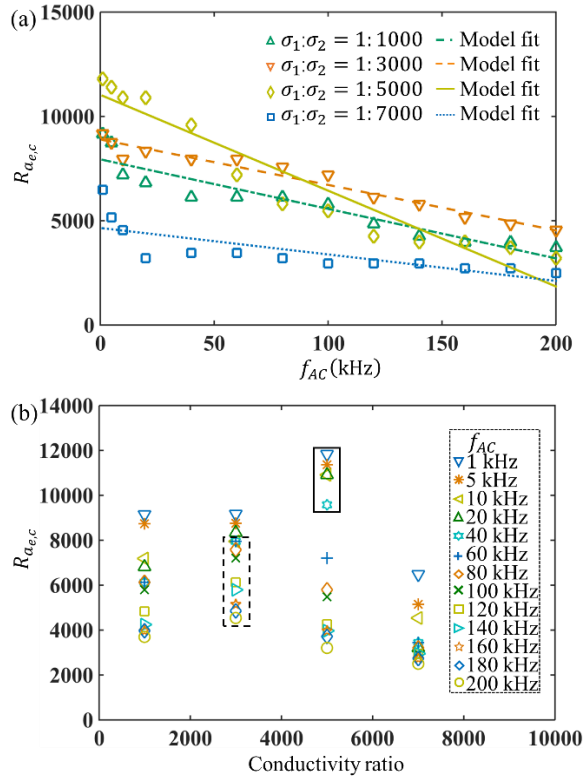

**Figure 7.** Relationship between $Ra_{e,c}$, $f_{AC}$ and conductivity ratio. (a) $Ra_{e,c}$ vs $f_{AC}$ at four different electric conductivity ratios. The points represent experimental results and the straight lines represent linear fitting results. (b) $Ra_{e,c}$ changes with $\sigma_1 : \sigma_2$ at different $f_{AC}$. The boxes indicate that when $f_{AC} \leq 40$ kHz, the highest $Ra_{e,c}$ appears at 1:5000 conductivity ratio, and the dotted line box indicates that when $f_{AC} > 40$ kHz, the highest $Ra_{e,c}$ appears at 1:3000 conductivity ratio.

## CONCLUSION

In this investigation, we systematically investigated the influence of electric field intensity, AC frequency, electric conductivity ratio, and channel width at the entrance on the mixing effect and transition electric Rayleigh number in the "Y" type electrokinetic micromixer. In a 350 μm wide micromixer, a maximum degree of mixedness of 0.93 can be realized at 84 μm distance from the entrance, under 100 kHz and $1.14 \times 10^5$ V/m AC electric field. To the best of our knowledge, this is the shortest distance that has ever seen. The time cost is only 100 ms.

A detailed investigation of the mixing region shows that mixing can be separated into two stages: a linear development region and a nonlinear saturation region. The streamwise positions of their interaction points generally show a power-law relationship with electric Rayleigh number. The higher the electric Rayleigh number, the closer the intersection point to the entrance. Thus, increasing the electric Rayleigh number can significantly enhance mixing and reduce the mixing length.

Further investigations on the critical electric field and the critical electric Rayleigh number indicate the most unstable condition of EK flow instability is inconsistent with that of the optimal mixing in EK turbulence. Although a smaller $Ra_{e,c}$ implies the electrokinetic flow can have stronger linear instability under a lower electric field, it is insufficient to induce a fast and uniform mixing under a higher electric field. To predict the evolution of EK flow under high $Ra_e$, it is necessary to apply a computational turbulence model, instead of linear instability analysis.

We hope the current electrokinetic micromixer can be developed as a general-purpose platform for industrial applications, with high throughput and high performance, in heat and mass transport, chemical analysis, reaction and synthesis, environmental monitoring etc.

**ACKNOWLEDGEMENT** This investigation is supported by National Natural Science Foundation of China (Grant No. 51927804, 61775181, 61378083).